\documentclass[aps,prl,twocolumn,showpacs,amsmath]{revtex4}
\usepackage{graphics}
\usepackage{epsfig}
\usepackage[colorlinks=true,citecolor=blue,linkcolor=red,linktocpage=true,pagebackref=false]{hyperref}
\usepackage{color,soul} 
\newcommand{\be}{\begin{eqnarray}}
\newcommand{\ee}{\end{eqnarray}}
\newcommand{\ben}{\begin{eqnarray*}}
\newcommand{\een}{\end{eqnarray*}}

\begin{document} 
   \title{
Glauber coherence of single electron sources
   }

\author{G. Haack$^{1,2}$, M. Moskalets$^{1,3}$, and M. B\"uttiker$^1$}
\affiliation{$^1$D\'epartement de Physique Th\'eorique, Universit\'e de Gen\`eve, CH-1211 Gen\`eve 4, Switzerland\\$^2$ Dahlem Center for Quantum Complex Systems and Fachbereich Physik, Freie Universit\"at Berlin, 14195 Berlin, Germany\\$^3$Department of Metal and Semiconductor Physics, NTU "Kharkiv Polytechnic Institute", 61002 Kharkiv, Ukraine
}
\date\today
 \begin{abstract}
Recently demonstrated solid state single electron sources generate different quantum states depending on their operation condition. For adiabatic and non-adiabatic sources we determine the Glauber correlation function in terms of the Floquet scattering matrix of the source. The correlation function provides full information on the shape of the state, on its time-dependent amplitude and phase, which makes the coherence properties of single electron states essential for the production of quantum multi-particle states.
\end{abstract}


\maketitle

\begin{figure}[t]
\includegraphics[width=8cm]{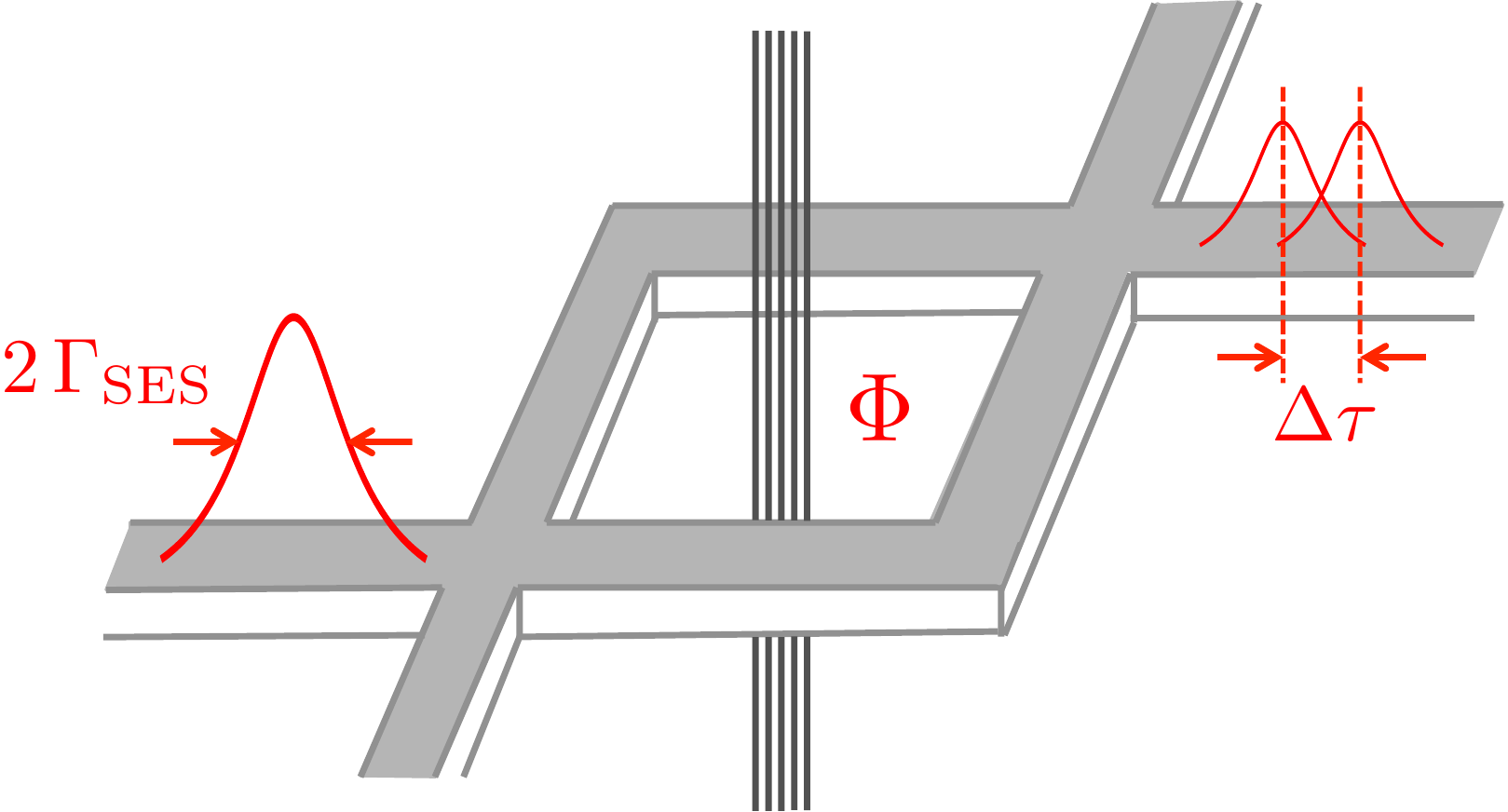}
\caption{Schematic representation of a MZI, threaded by a magnetic flux $\Phi$. With a time-resolved measurement of the current in one of the output arms, one can access the first-order correlation function $G^{(1)}$ as a function of the time delay of the interferometer $\Delta \tau$ and time $t$. This allows to reconstruct the incoming single-particle state emitted by the SES. In the adiabatic regime, the current pulse emitted by the SES has a Lorentzian shape, with a width $2 \Gamma_\mathrm{SES}$.}
\label{fig1}
\end{figure}

{\textit{Introduction --}} 
The recent realization of triggered electron sources that inject single electrons on demand into high mobility semiconductors attracts increasing attention to the field of quantum coherent electronics \cite{Feve07, Blumenthal07, Splettstoesser09, Bocquillon12, Kashcheyevs12}. Future applications in quantum information processing demand a full characterization of the coherence of the states emitted by such sources \cite{Grenier11,Haack11}. The important feature of on-demand injected particles is that they are traveling wave-packets with a spatial extend that is less than the distance between them. Depending on the operating conditions of the source, wave-packets of different spatial and temporal shape can be created \cite{Feve07, Bocquillon12}. Such wave packets are able to interfere with themselves over a restricted interval of space and time, which sets the limits on the synchronization of multiple single electron sources needed to generate on demand multi-particle states. It is the purpose of this work to present a full characterization of the coherence of the single particle states generated by on-demand sources. 
\\ \indent
In optics the coherence of light is discussed with the help of correlation functions introduced by Glauber \cite{Glauber63}. The first-order correlation function reads, $G^{(1)}\left( r_{1}t_{1}, r_{2}t_{2} \right) = \left\langle  E^{(-)}\left( r_{1}t_{1} \right) E^{(+)}\left( r_{2}t_{2} \right) \right\rangle$,  where the electric field of a light-beam is split into positive  $E^{(+)}$ and negative $E^{(-)}$ frequency terms \cite{Glauber06}. 
The first-order Glauber correlation function can be extracted from time- and space-resolved intensity (optics) or current (electronic) at the output of an interferometer, see Fig.~\ref{fig1}. Remarkably the characterization of single photons has been achieved very recently with space-resolved measurement of the intensity \cite{Lundeen11, Polycarpou12}. In mesoscopic systems, time-resolved current measurements on the scale of single electron wave packets have recently been demonstrated \cite{Feve07}. This makes it possible to reconstruct the single-particle state from current measurements, as well as the complex wave function, the duration of the wave packet and the coherence time. Therefore the Glauber correlation function is the central object and, in this Letter, we discuss it for the states of adiabatic and non-adiabatic emitters. Importantly, for a single-particle state the second and higher-order correlation functions are zero since not more than one particle can be measured at a time \cite{Lounis05}. For a source that emits particles periodically, the second order correlation function is measured to demonstrate a single-photon source \cite{Grangier91,Nguyen:2011dh}. By analogy the single-particle nature of an electron state of interest here can be inferred from the zero frequency current noise measurement \cite{Bocquillon12}. 

The fermionic first-order correlation function can be defined in close analogy with the bosonic one \cite{Cahill99}. However the single electrons we are interested in are injected into the conductor with other electrons constituting the Fermi sea. Importantly the underlying Fermi sea has a non-zero correlation function which can be naturally treated as the reference point \cite{Grenier11}. Therefore, we define the first-order correlation function as $G^{(1)}\left(t_1,t_2\right) =\langle \hat\Psi^{\dag}\left(t_1\right) \hat\Psi\left(t_2 \right) \rangle - \langle \hat\Psi^{\dag}\left(t_1 \right) \hat\Psi\left(t_2\right) \rangle_{0}$,  with $\hat\Psi\left( t_{1,2} \right)$ a single-particle electronic field operator at times $t_{1,2}$. We omit the spatial coordinates $(r_1,r_2)$ of the correlation function, since the current is measured in the reservoir at $\!r_{1}\!\!=\!\! r_{2}$. The angular brackets denote the quantum-statistical average over the state of the Fermi sea and the subscript $0$ indicates that the single-electron source (SES) is not active. This electronic first-order correlation function $G^{(1)}$ is accessible in a Mach-Zehnder interferometric (MZI) set-up. The electronic MZI was first reported in the two-dimensional electron gas in high magnetic field in the quantum Hall regime \cite{Ji03}. Experimentally it has been shown to exhibit high visibility while varying a phase $\phi$ by tuning the magnetic flux $\Phi$ enclosed by the arms of the MZI  and/or  the time delay between its arms. Below we show that the interference part of the current at the output of the MZI is written in terms of the correlation function $G^{(1)}$ as follows,
\begin{eqnarray}
I_{out}^{int}(t) \propto {\rm Re} \left\{ e^{- i \phi} G^{(1)}(t- \tau_\mathrm{u}, t- \tau_\mathrm{d}) \right\} \,.
\label{eq01}
\end{eqnarray}
Here $\tau_\mathrm{u,d}$ are the traversal times for the upper and lower arms of the MZI. Fixing the phase $\phi$ to zero or $\pi/2$ gives access experimentally to the real or imaginary parts of the correlation function respectively. 
This allows us to extract the shape of the single-particle state, its phase and its coherence properties from a measurement of the full time-dependence of the first-order correlation function. 
The most challenging step, the time-resolved measurement of a current at a nano-second scale characteristic for a single-electron wave-packet, was recently shown to be possible \cite{Feve07}.   

{\textit{Model and first-order correlation function --}} To be specific we focus on single-particle states emitted by the on-demand source  of Ref.~\onlinecite{Feve07}. This source consists of a mesoscopic capacitor \cite{Buttiker94, Gabelli06,Nigg:2006kl,Mora:2010hw} driven by a periodic potential $V(t)$. 
Built in the quantum Hall regime, the SES is made of a small cavity with a confined circular edge state, which is connected via a quantum point contact (QPC) with transmission $T_\mathrm{SES} \ll 1$ to the nearby linear edge state. By shifting the levels of the cavity above and below the Fermi sea level with $V(t)$, the emission of a single electron and a single hole in one period of the potential is achieved \cite{Feve07}. 
Within a scattering-matrix approach, the SES is described by a Floquet scattering amplitude $S_\mathrm{SES}(E_m, E)$, calculated in Ref.~\onlinecite{Moskalets08}, where the energy of the outgoing particle $E_m = E + m\hbar \Omega$ differs from the energy $E$ of the incoming particle by $m\hbar \Omega$. Here $\Omega$ is the frequency of the periodic potential and $m$ is an integer. In the quantum Hall regime, the chirality of the edge states due to the absence of backscattering \cite{Halperin82, Buttiker88} allows us to write the scattering amplitude of the entire system $S(E_m,E)$ as the product of the scattering amplitude of the MZI, calculated at energy $E_m$, with the Floquet scattering amplitude $S_\mathrm{SES}(E_m,E)$ of the source \cite{Splettstoesser09, Haack11}. Then the outgoing current is expressed in terms of a current emitted by the cavity, $I_\mathrm{SES}$, and of the first-order correlation function introduced above, $G^{(1)}$:
\begin{eqnarray}
I_{out}(t)  = R_\mathrm{L}R_\mathrm{R}\, I_\mathrm{SES} (t- \tau_\mathrm{u}) + \, T_\mathrm{L}T_\mathrm{R}\, I_\mathrm{SES} (t- \tau_\mathrm{d})\nonumber  
\\
-  2 \sqrt{R_\mathrm{L}R_\mathrm{R}T_\mathrm{L}T_\mathrm{R}} \, e v_{D}\,\mathrm{Re}\Big\{ e^{-i\phi} G^{(1)} (t- \tau_\mathrm{u},t- \tau_\mathrm{d}) \Big\} . 
\label{eq:main_results1}
\end{eqnarray}
The coefficients $R_\mathrm{L,R}$ and $T_\mathrm{L,R}$ are the reflection and transmission probabilities for the left and right QPCs of the MZI respectively. The term $\phi = 2\pi \Phi/\Phi_0 + k_\mu v_D \Delta \tau$ corresponds to the phase difference acquired by an electron with Fermi energy $\mu$ traveling along the upper and lower arms of interferometer, where $\Phi_0=h/e$ is the quantum flux, $k_\mu$ and $v_{D}$ are the wave vector and the drift velocity both evaluated at the Fermi energy and $\Delta \tau = \tau_\mathrm{u} - \tau_\mathrm{d}$ is the time-delay of the interferometer. The time-dependent current emitted by the source is \cite{Splettstoesser08}
\be
&&I_\mathrm{SES}(t) = \frac{e}{h}\int_{0}^\infty dE\sum_m\left[f(E)-f(E_m)\right] \nonumber \\
&&\ \ \ \int\frac{dt'}{\mathcal{T}}e^{-im\Omega(t'-t)}S_\mathrm{SES}^*(t',E)S_\mathrm{SES}(t,E)\,,
\label{eq:current_cavity}
\ee
and the first-order correlation function is expressed in terms of the Floquet scattering amplitude of the source $S_\mathrm{SES}$ as follows (we denote $t_\mathrm{u}\equiv t- \tau_\mathrm{u}$ and $t_\mathrm{d}\equiv t- \tau_\mathrm{d}$):
\be
G^{(1)} (t_\mathrm{u},t_\mathrm{d}) = \int_{0}^{\infty}\frac{dE}{hv_{D}} \sum_{m} \left[f(E)-f(E_m)\right]  \ \ \ 
\label{eq:main_results2}
 \\
 e^{-i (E-\mu) \frac{\Delta \tau}{\hbar} }\!\!\! \int \frac{dt'}{\mathcal{T}} e^{-im \Omega(t'-t_\mathrm{u})} \,S^*_{\mathrm{SES}}(t' ,E) S_\mathrm{SES}(t_\mathrm{d},E)\,.
 \nonumber
\ee
Here we have introduced the Floquet scattering amplitude of the source in a mixed energy-time representation, $S_\mathrm{SES}(t,E) = \sum_{n} e^{-in\Omega t} S_\mathrm{SES}\left( E_{n}, E \right)$. Importantly Eq.(\ref{eq:main_results2}) derived here is valid at arbitrary emission conditions. 
This is in contrast to Ref.\cite{Haack11} where we used the version of Eq.(\ref{eq:main_results2}) valid in the adiabatic regime only. Moreover, in Ref.\cite{Haack11}, we defined the single-particle coherence on the basis of an interference current. In contrast, in the present Letter, we adapt  the Glauber definition of the correlation function and show precisely how it is connected to the interference current.

\begin{figure}[t]
\includegraphics[width=9cm]{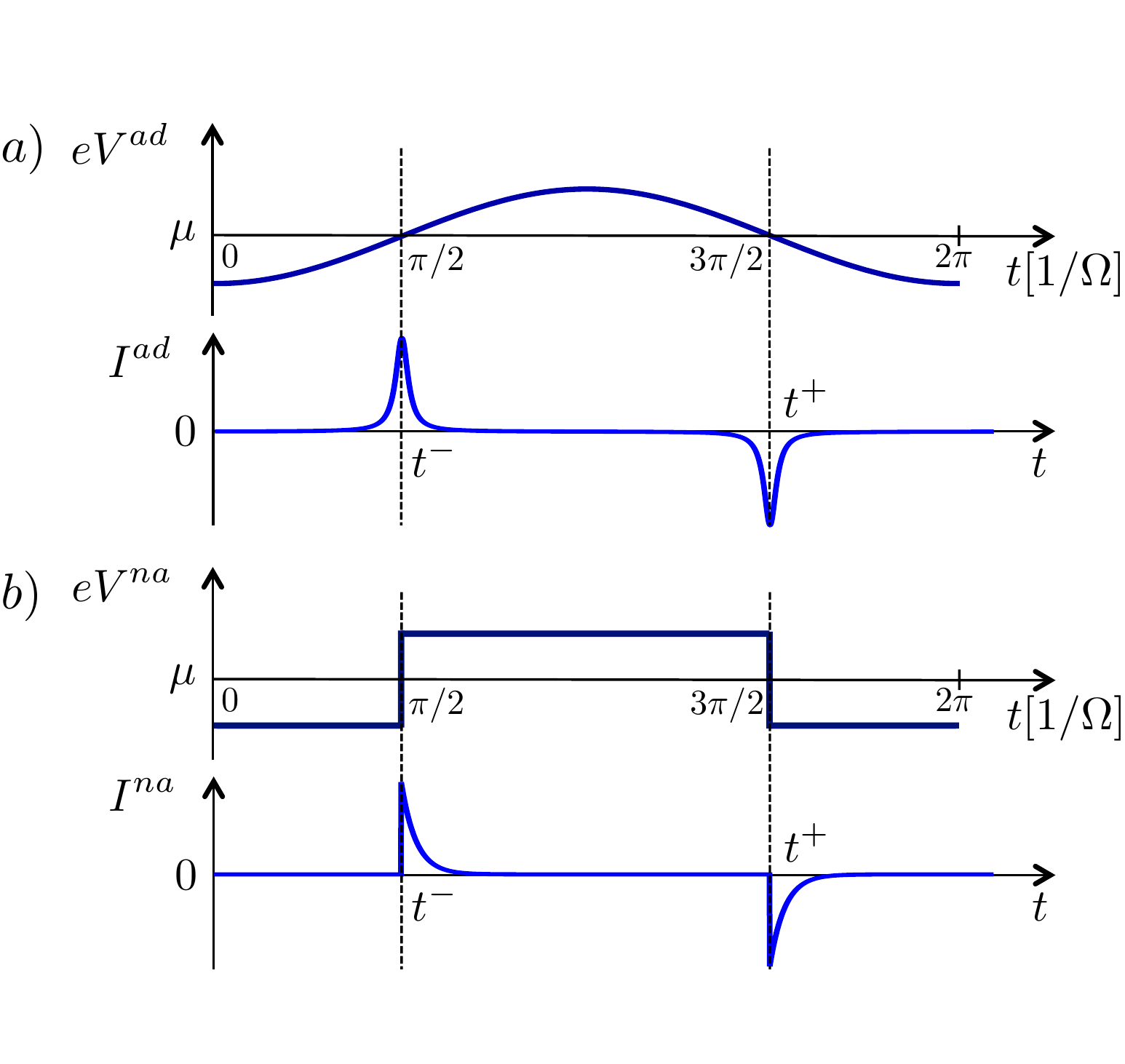}
\caption{Time-dependent potential $V$ driving the SES and induced current $I$ consisting of electron and hole pulses are shown for one period of $V$: a) in the adiabatic regime and b) in the non-adiabatic regime. The emission process takes place when $V$ makes the topmost occupied level cross the Fermi energy $\mu$. The emission times of an electron and a hole are respectively denoted $t^-=\pi/2 \Omega$ and $t^+=3\pi/2 \Omega$. The strong symmetric/asymmetric shape of the current pulse is characteristic of the adiabatic/non-adiabatic emission process.}
\label{fig2}
\end{figure}

\begin{figure}[t]
\includegraphics[width=7cm]{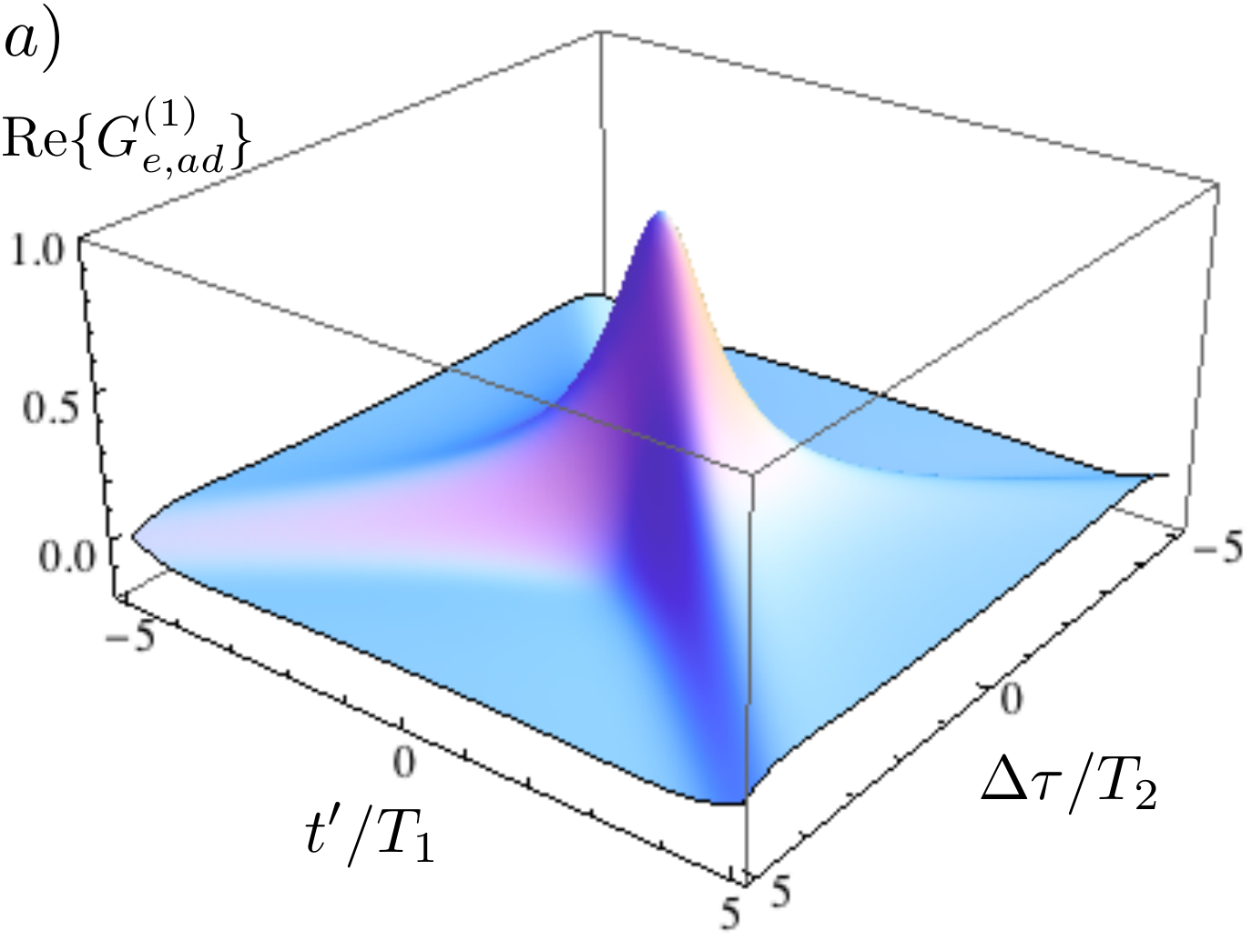}
\includegraphics[width=7cm]{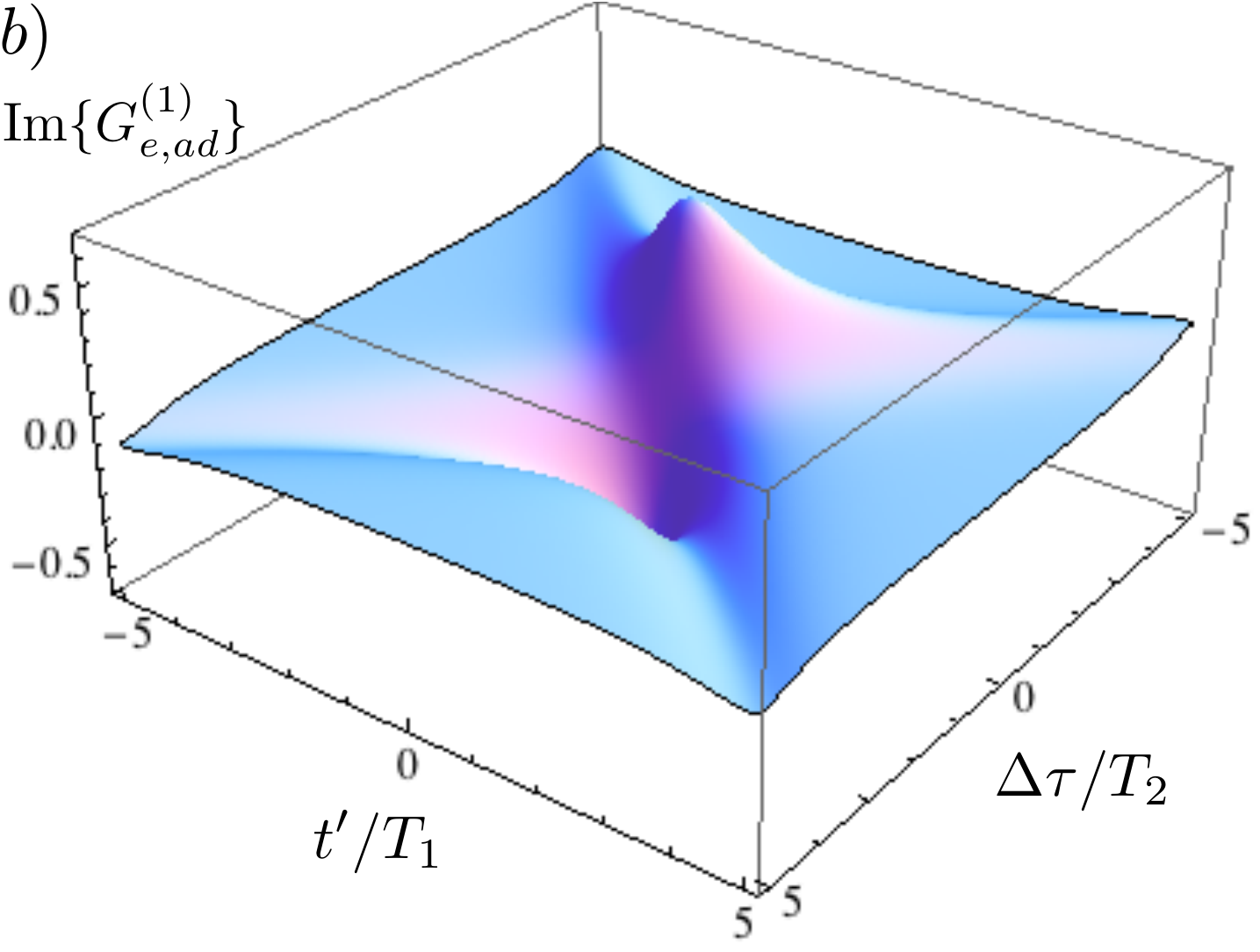}
\caption{Real (a) and imaginary (b) part of the first-order correlation function $G^{(1)}_{e,ad}(t' - \Delta \tau/2, t' + \Delta \tau/2)$ for electrons emitted adiabatically (Eq.(\ref{eq:coherence}) in units of $1/(\pi \Gamma v_{D})$). Here we set $t^-\equiv 0$. At $t'= \Delta \tau=0$, the overlap of the wave-packets is maximal ($G^{(1)}_{e,ad}=1$ in normalized units). The decrease of $G^{(1)}_{e,ad}$ as a function of $t'$ and $\Delta \tau$ is set respectively by the lifetime $T_1 = \Gamma$ and the coherence time $T_2=2 \Gamma$.  The real part of the $G^{(1)}_{e,ad}$ function at $\Delta \tau = 0$ corresponds to the current pulse emitted by the source as a function of $t'$, whereas its imaginary part is zero as expected.}
\label{coh-ad}
\end{figure}

{\textit{Adiabatic versus non-adiabatic regimes --}} 
We illustrate our claim that we can fully characterize the single-particle state by its first-order correlation function, Eq.(\ref{eq:main_results2}), by considering the source of Ref.~\cite{Feve07} in the two operation regimes in which single-particle emission can be achieved, namely the adiabatic and non-adiabatic regimes. In the following, we assume zero temperature. If the temporal shape of the periodic driving potential $V(t)=V( t + 2\pi/\Omega)$ varies on a time scale much smaller than the dwell time $\tau_{D}$ of the source, defined as the time that the particle remains inside the cavity, the operation regime of the source is called adiabatic \cite{Moskalets02}. Experimentally, it can be reached with a sinusoidal potential $V^{ad}(t) = V_{0}\cos(\Omega t)$ with $\Omega\tau_{D} \ll T_\mathrm{SES}$ \cite{Splettstoesser08}. 
This last assumption ensures that an electron has enough time to leave the cavity during the time when the topmost occupied level crosses the Fermi energy., see Fig.~\ref{fig2} a).
Here $V_0$ is the amplitude of the potential. In this regime, the single-particle states are emitted close to the Fermi sea and the energy in Eqs.(\ref{eq:current_cavity},\ref{eq:main_results2}) is therefore well approximated by the Fermi energy $\mu$. The SES is described by the frozen scattering amplitude \cite{Moskalets08}, which, close to the emission time $t^{-}$ of an electron, reads \cite{Olkhovskaya08}: $S_{\mathrm{SES},e}^{ad}\left( t,\mu \right) \!=\! \left( t - t^{-} + i \Gamma \right)\!/\!\left( t - t^{-} - i \Gamma \right)$. The corresponding current emitted by the SES consists of a Lorentzian pulse, $I_{e}^{ad}\left( t \right) \!=\! (e\Gamma/\pi)/\!\left( \left[t - t^{-} \right]^{2} + \Gamma^{2}\right)$, where the half-width of the current pulse $\Gamma$ is proportional to $T_\mathrm{SES}/\Omega$. Importantly, it sets the lifetime (or the relaxation time) of the emitted single-particle state, $T_{1}^{ad} = \Gamma$. 
To find the coherence time of the emitted state $T_{2}$ we look at the correlation function  which now reads:
\be
G^{(1)}_{e,ad} (t_\mathrm{u},t_\mathrm{d}) = \frac{1}{\pi \Gamma v_{D} } \frac{1}{\left(1 - i\frac{ t_\mathrm{u} - t^-}{\Gamma } \right)\!\left(1 + i\frac{ t_\mathrm{d} - t^-}{\Gamma } \right) }.
\label{eq:coherence}
\ee
The characteristic time of decay of $G^{(1)}$ with respect to the time delay $\Delta \tau\! =\! \tau_\mathrm{u} \!-\! \tau_\mathrm{d}$ is by definition the coherence time $T_2$ of the single-particle states. 
To make clear the dependence on $\Delta \tau$ we introduce the middle time $t' = (t_\mathrm{u} +t_\mathrm{d}) /2$ and write $t_\mathrm{u} = t' - \Delta \tau/2$, $t_\mathrm{d} = t' + \Delta \tau/2$.  
Thus we find from Eq.~(\ref{eq:coherence}) that $T_2$ is set by twice the lifetime of the current pulse, $T_2^{ad} = 2 \Gamma$. 
The relation $T_{2}^{ad} = 2 T_1^{ad}$ means that the emitted state is a Fourier-transform limited one \cite{Devoret04}. This important result tells us 
that the SES has no intrinsic dephasing time $T_\varphi$, since the three times are related via $1/T_2 = 1/(2T_1) + 1/T_\varphi$ \cite{Abragam85, Lounis05}. Additional dephasing processes within the MZI  would lead to a faster decay of the interference part of the measured current \cite{Roulleau08,Bieri09,Altimiras10,Sueur10}, but would not modify the coherence properties of the states emitted by the source. The real and imaginary parts of the correlation function for adiabatically emitted electrons are shown in Fig.~\ref{coh-ad}. They allow to reconstruct the shape of the incoming wave-packet as well as its phase \cite{Kano62}. The correlation function for the hole, $G_{h,ad}^{(1)}$, is given by the complex conjugate of  Eq.(\ref{eq:coherence}), where the electron emission time $t^{-}$ is replaced by the hole emission time $t^{+}$.  

\begin{figure}[t]
\includegraphics[width=7cm]{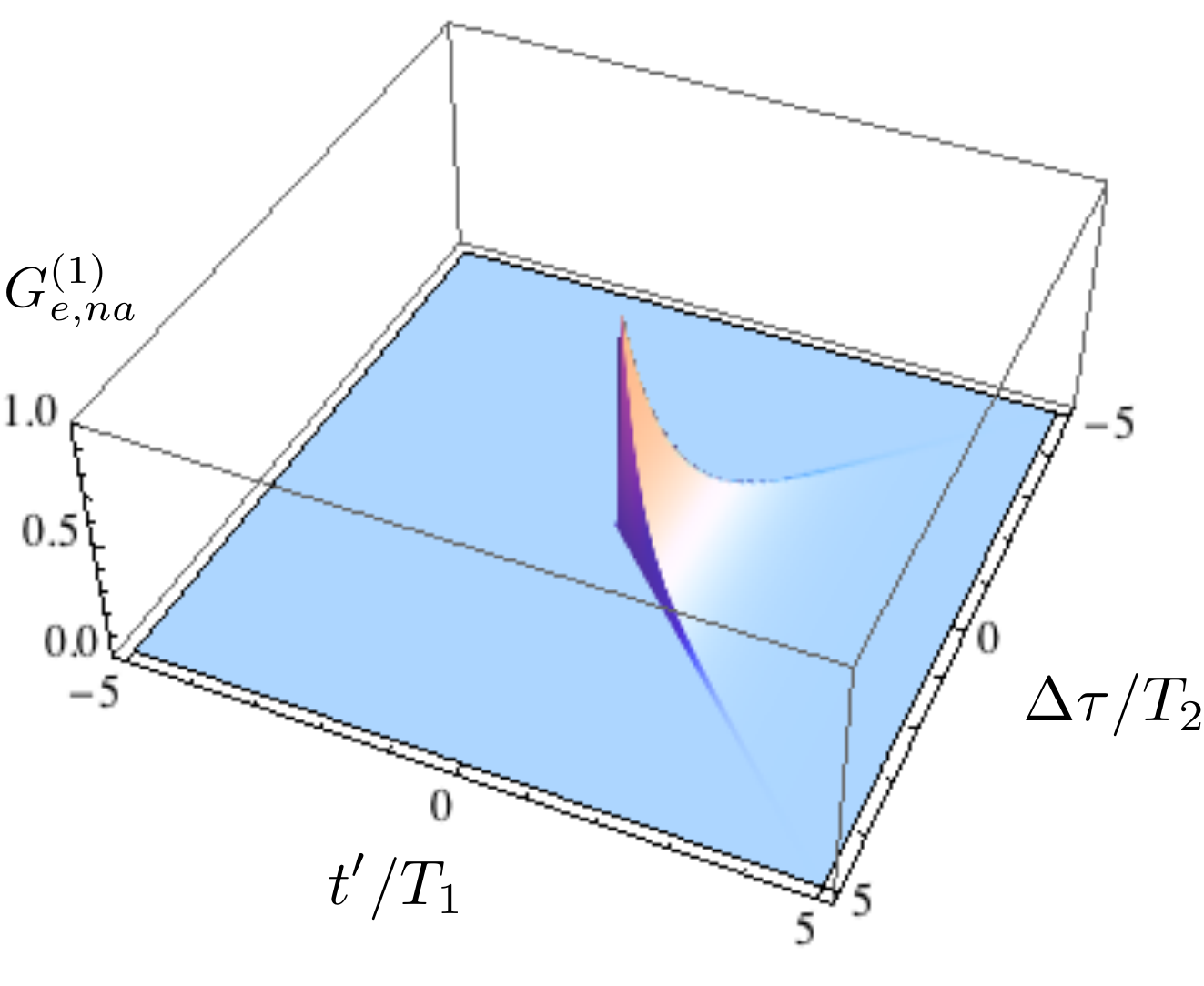}
\caption{First-order correlation function for electrons emitted non-adiabatically, $G^{(1)}_{e,na}(t' - \Delta \tau/2, t' + \Delta \tau/2)$, Eq.~(\ref{cohna}), in units of $1/(\tau_{D} v_{D})$. The exponential factor $e^{-i\pi \Delta\tau/\tau}$ is omitted as it sets the energy at which the single-particle state is emitted (see text). Here $t^-$ is set to 0. The correlation function clearly reflects the temporal shape of the single-electronic state emitted by the source, which is set by $T_1$ and $T_2$ as a function of $t'$ and $\Delta \tau$ respectively.}
\label{fig4}
\end{figure}

The non-adiabatic regime is reached when the driving potential varies much faster than the dwell time $\tau_{D}$. Experimentally the emission of single-particle states has been observed in this regime with a square potential in the GHz range \cite{Feve07}. Importantly, while the potential changes on a time scale faster than $\tau_{D}$, the overall cycle remains much longer than $\tau_{D}$, ensuring that an electron has been emitted before the excitation leading to the hole emission 
starts, see Fig. \ref{fig2} b), \cite{Parmentier12}.
This corresponds to the condition $\tau_D \ll \pi/\Omega$, which can be fulfilled at higher frequencies than the condition for an adiabatic regime. To provide simple analytical equations we assume the optimal conditions used in the experiment \cite{Feve07,Mahe10}: the Fermi level lies exactly in the middle of two successive cavity's levels and the square potential $V^{na}(t)$ applied to the cavity shifts the levels sharply by one level spacing $\Delta$ at time $t^{-}$. With such a driving, the Floquet amplitude given in Ref.~\onlinecite{Moskalets08} can be cast into a form appropriated for analytical calculations:
\be
S_\mathrm{SES}^{na}(E_{n},E) &=& S(E)  e^{ i \pi \frac{n \hbar\Omega  }{\Delta }  }\,\frac{\sin\left( \pi n \frac{ \hbar\Omega }{\Delta } \right) }{\pi n } 
\label{flna} \\
&&\times \left\{  \frac{\Delta }{\hbar\Omega } \delta_{n,0} - \frac{ \frac{ e^{i n \Omega t^{-} } }{1 -  \frac{n\hbar\Omega }{\Delta }  } + \frac{ e^{i n \Omega t^{+} } }{1 +  \frac{n\hbar\Omega }{\Delta } }  }{\rho^{*}\left( E \right) \rho\left( E_{n} \right) }
 \right\}
 \,. \nonumber 
\ee
Here $\rho(E) \!=\!\big(\!1 \!+\! \sqrt{1\!-\!T_\mathrm{SES}}\! \exp( \!i\phi(E)) \!\big)\!/\!\sqrt{T_\mathrm{SES}}$ with $\phi(E)= 2\pi( \!E \!-\! \mu)\!/\!\Delta $ and $S(E) \!=\! \exp\!\big( i\phi(E) \big) \rho^*(E)/\rho(E)$ is the  scattering amplitude of the cavity with stationary potential. Since $\tau_{D} = h/(T_\mathrm{SES} \Delta ) \ll 2\pi/\Omega$, the emissions of an electron and a hole close to $t^-$ and $t^+$ are independent of each other. Therefore, as before, we concentrate on electron emission only.
Calculating the current emitted by the SES close to $t^{-}$ from Eq.(\ref{eq:current_cavity}), we reproduce a well-known exponential decay \cite{Feve07, Moskalets08}, $I_{e}^{na}(t) = (e/\tau_{D}) \Theta(t - t^{-}) \, e^{-(t -t^{-})/\tau_{D}}$ with $\Theta(x)$ the Heaviside step function. From the temporal shape of the current pulse, we extract the lifetime of the single-particle state in the non-adiabatic regime, namely $T_{1}^{na} = \tau_{D}$. Remarkably, in contrast to the current pulse in the adiabatic regime, the pulse $I_{e}^{na}(t)$ is highly asymmetric in time as shown in Fig.~\ref{fig2} b) \cite{Keeling08, Battista12}. This strong asymmetry is a signature of a non-adiabatic emission process and is also present in the first-order correlation function. Indeed, inserting Eq.(\ref{flna}) into Eq.(\ref{eq:main_results2}) we find:
\be
G^{(1)}_{e,na}(t_\mathrm{u},t_\mathrm{d}) =  \frac{1}{\tau_{D} v_{D}} \, \Theta(t_\mathrm{u} - t^{-})\Theta(t_\mathrm{d} - t^{-}) 
\label{cohna} \\
\times \exp\left(-i \pi\frac{  \Delta\tau}{\tau } \right) \exp\left(  -\frac{(t_\mathrm{u} + t_\mathrm{d})/2  - t^{-}}{ \tau_{D} } \right)    .    
\nonumber 
\ee
The factor $\exp\left(-i \pi \Delta\tau/\tau \right)$ reflects the fact that the single-particle states are emitted at energy $\Delta/2$ above the Fermi energy $\mu$, ($\tau \equiv h/\Delta$). Due to the presence of the Heaviside step functions,  the middle time $t' \!=\! (t_\mathrm{u} \!+\!t_\mathrm{d}) /2$ has to be larger than $t^{-} + \Delta\tau/2$ for $G^{(1)}_{e,na}$ to be non-zero, as shown in Fig.~\ref{fig4}. Thus we see that the first-order correlation function decays with increasing $\Delta \tau$ with a characteristic time $T_{2}^{na} = 2\tau_{D}$.  
Similarly to the adiabatic regime, the coherence time is equal to twice the lifetime, $T_{2}^{na}  = 2T_{1}^{na}$, witnessing the absence of intrinsic dephasing in the SES. 

{\textit{Conclusions --}} We have shown that an MZI setup is appropriate for the full characterization of the coherence properties of single electrons and holes propagating in solids. 
We have provided a general expression for the Glauber correlation function $G^{(1)}$ in terms of the Floquet scattering amplitude of the source. 
The coherence time enabled us to show that the source of Ref.~\onlinecite{Feve07} 
has no intrinsic dephasing time, which makes the emitted single-particle states of high interest for future experiments in quantum electronics. 
Importantly, the time-resolved measurement of the first-order correlation function $G^{(1)}$ is within the reach of the present-day experimental capabilities, permitting a direct access to a single-electronic quantum state.

{\textit{Acknowledgements --}} G.H. thanks A. Baas and M. Richard for clarifying discussions on the first-order correlation function and its measurement in quantum optics. This work was supported by the Swiss NSF and the Swiss NCCR on Quantum Science and Technology QSIT and G.H. also acknowledges support from the Alexander von Humboldt Foundation.

\end{document}